\documentclass[a4paper,10pt,twoside]{cpc-hepnp}
\usepackage{CJK,upgreek,fancyhdr}
\usepackage{multicol}
\usepackage{graphicx}
\usepackage{booktabs}
\usepackage{amssymb,bm,mathrsfs,bbm,amscd}
\usepackage[tbtags]{amsmath}
\usepackage{lastpage}

\newcommand{\be}{\begin{equation}}
\newcommand{\ee}{\end{equation}}
 \newcommand{\bea}{\begin{eqnarray}}
 \newcommand{\ena}{\end{eqnarray}}

\begin{document}
\begin{CJK*}{GB}{gbsn}

\fancyhead[c]{\small Chinese Physics C~~~Vol. xx, No. x (201x) xxxxxx}
\fancyfoot[C]{\small 010201-\thepage}

\footnotetext[0]{Received   }

\title{Thermodynamics and weak cosmic censorship conjecture of the BTZ black holes in the  extended phase space\thanks{Supported by National Natural Science
Foundation of China(Grant Nos. 11875095), and Basic Research Project of Science and Technology Committee of Chongqing (Grant No. cstc2018jcyjA2480, cstc2019jcyjA2091). }}

\author{%
      Xiao-Xiong Zeng(ÔøÏþÐÛ)$^{1;1)}$\email{xxzengphysics@163.com}%
\quad Yi-Wen Han (º«ÒàÎÄ)$^{2}$
\quad De-You Che(³ÂµÂÓÑ)$^{3}$
}
\maketitle

\address{%
$^1$ Department of Mechanics, Chongqing Jiaotong University, Chongqing 400074, China \\
$^2$ School of Computer Science and Information Engineering, Chongqing Technology and Business University,\\ Chongqing 400070, China\\
$^3$School of Science, Xihua University, Chengdu 610039, Sichuan, China\\
}

\begin{abstract}

As  a charged fermion drop into a BTZ black hole,   the laws of thermodynamics and the weak cosmic censorship conjecture are checked in both the normal phase space and extended phase space, where the cosmological parameter and renormalization length are regarded as  extensive quantities. In the normal phase space, the first  law, second law, and the weak cosmic censorship are valid. While in the extended phase space, though the first law and weak
cosmic censorship conjecture   are  still  valid, the second law  is dependent  on the variation of the renormalization energy
$dK$. In addition, in the  extended phase space, the  configurations  of  the extremal and near-extremal black holes will not be changed  for they are stable while   in  the normal phase space,  the extremal and near-extremal  black holes will evolve into non-extremal black holes.
\end{abstract}



\footnotetext[0]{\hspace*{-3mm}\raisebox{0.3ex}{$\scriptstyle\copyright$}2013
Chinese Physical Society and the Institute of High Energy Physics
of the Chinese Academy of Sciences and the Institute
of Modern Physics of the Chinese Academy of Sciences and IOP Publishing Ltd}%

\begin{multicols}{2}

\section{Introduction}

\label{intro}
From the pioneering work of Hawking \cite{Hawking:1974sw,Hawking:1976de}, we now know that a black hole can be regarded as a thermodynamic system. Similar to the   usual thermodynamic systems, there are four laws of thermodynamics for the black holes. The event horizons of the black holes play a key role in the thermodynamic systems for both the temperature and entropy are related to it.
In addition, the event horizon will hidden the singularity of the spacetime,   otherwise the  weak cosmic censorship conjecture proposed by  Penrose \cite{Penrose:1969pc} will be violated. The Kretschmann scalar can be used to check  the  weak cosmic censorship conjecture since it is  independent of
the choice of coordinates \cite{Henry:1999rm,Garfinkle:1998va,Ziaie:2011dh}. The location where the  Kretschmann scalar
is infinite is the singularity.

The laws of thermodynamics and weak cosmic censorship conjecture can be  checked with consideration of   a test particle \cite{Wald:1974ge,Hubeny:1998ga,Gwak:2016gwj,Jacobson:2009kt,Colleoni:2015afa,Hod:2016hqx,Gwak:2017icn,Gao:2012ca,Rocha:2011wp,Sorce:2017dst,Ge:2017vun,An:2017phb} or a test field \cite{Semiz:2005gs,Toth:2011ab,Duztas:2016xfg,Natario:2016bay,Chen:2018yah,Chen:2019nsr,Gwak:2018akg,Gwak:2019asi}. In \cite{Gwak:2015fsa}, the first law, second law as well as the weak cosmic censorship conjecture of a BTZ black hole have been investigated. It was found that the first and second laws were valid and the  weak cosmic censorship conjecture held for the extremal black hole. Later, the idea in \cite{Gwak:2015fsa} was extended to  a $D$-dimensional charged AdS black hole in the extended phase space, where the negative cosmological constant and its conjugate  were regarded as the pressure and volume respectively \cite{Gwak:2017kkt}.  An interesting result in \cite{Gwak:2017kkt} is that the second law is violated for the extremal and near-extremal black holes as the contribution of the pressure and volume are considered. In addition, the extremal black holes are found to be stable for the absorbed particles will not change the configurations of the black holes. Recently, \cite{Zeng:2019jta} and \cite{han} investigated thermodynamics and weak cosmic censorship conjecture in the  Born-Infeld AdS black holes   and phantom Reissner-Nordstr\"{o}m  AdS black holes. Different from the  result in \cite{Gwak:2017kkt}, they found that the extremal black holes will change into non-extremal black holes. The reason stems from that they did not employ any approximation while \cite{Gwak:2017kkt} employed.

It should be noted that in the works mentioned above, they considered only the case that the black holes absorb  scalar particles. In this paper, we will study the case of fermions with the Dirac equation. We intend to explore whether we can obtain the same result for the result is not priori. We will take the  BTZ black holes as an example.

There are two viewpoints on the thermodynamics of the BTZ black holes in the extended phase space. Identifying the cosmological constant and its conjugate   as the thermodynamic pressure and thermodynamic volume, the first law  was derived directly in \cite{Gunasekaran}. However, in their treatment, the   Reverse Isoperimetric Inequality is violated and the
black holes are always superentropic. Moreover, the thermodynamic volume defined by the first law is related to the charge. Soon after, \cite{Frassino:2015oca} introduced a new extensive
quantity, namely the renormalization
length, in the first law. In this framework, the   Reverse Isoperimetric Inequality is satisfied and the standard definition of the thermodynamic
volume is retained. In this paper, we will check whether the first law proposed by \cite{Frassino:2015oca} can be reproduced under a charged fermion absorbtion. Besides the first law, we will also investigate the second law as well as  the  weak cosmic censorship conjecture. As a result, the first law and
the  weak cosmic censorship conjecture are found to be valid while the second law is found to be related to the variation of the renormalization energy $dK$.

This paper is outlined as follows. In section 2  the motion of a charged fermion in the BTZ black holes is investigated. In section~3 and section~4,
 the first law, second law as well as  weak cosmic censorship conjecture are investigated in the normal phase space and extended phase space respectively. Section~5 is devoted to our conclusions.
Throughout this paper, we will set  $G=c=1$.

\section{Motion of a charged fermion in the BTZ black holes}

\label{secbh}

The three dimensional theory of gravity with Maxwell tensor  is \cite{BTZ1}
\begin{equation}
I=\frac{1}{16\pi G}\int d^{3}x\sqrt{-g}\left( R-2\Lambda -4\pi GF_{\mu \nu
}F^{\mu \nu }\right) ,~~~~~  \label{Einstein_Maxwell}
\end{equation}%
in which $G$ is the gravitational constant, $R$ is the Ricci scalar, $g$ is determinant of the
metric tensor $g_{\mu \nu }$, $\Lambda$ is the
cosmological constant that relates to the AdS radius with the relation $\Lambda =-1/{l^{2}}$, and  $F_{\mu \nu }=A_{\nu ,\mu }-A_{\mu ,\nu }$, where $A_{\mu }$  the electrical potential. The charged BTZ black hole solutions  can be derived from Eq.(\ref{Einstein_Maxwell}), that is
\begin{equation}
ds^{2}=-f\left( r\right) dt^{2}+f^{-1}\left( r\right) dr^{2}+r^{2}d\phi ^{2},
\label{bh_Q}
\end{equation}
where
\begin{equation}
f(r)=- {m}+\frac{r^{2}}{l^{2}}-2q^{2}\log  \left( \frac{r}{l}\right),
\label{frQ}
\end{equation}
in which $m$ and $q$ are the  parameters   which relate to the mass and charge of the black hole.
The non-vanishing component of the vector potential of this black hole is \cite{BTZ1}
\begin{equation}
A_{t}=q\log  (\frac{r}{l}).
\end{equation}%
Using the Gauss law, the electric charge of the black hole can be obtained by calculating the flux of the electric field at infinity \cite{Hendi:2016pvx}, which yields
\begin{equation}
Q= \frac{q}{2}. \label{charge}
\end{equation}%
In addition, the total mass can be got by using the Hamiltonian approach  or the counterterm method \cite{Hendi:2016pvx}, which leads to
\begin{equation}
M= \frac{m}{8}. \label{mass}
\end{equation}%

 Now we turn to investigate the  dynamical of a charged  fermion as it is absorbed by the BTZ black hole. We will employ the Dirac equation for
electromagnetic field
\begin{equation}
i\gamma ^{\mu }\left( \partial _{\mu }+\Omega _{\mu }-\frac{i}{\hbar }%
eA_{\mu }\right) \psi -\frac{\mu }{\hbar }\psi =0, \label{dirac}
\end{equation}
in which $u$ is the rest mass, $e$ is the charge of the  fermions, $\Omega _\mu = \frac{i}{2}\Gamma _\mu ^{\alpha \beta
} \sum _{\alpha \beta } $, $\sum _{\alpha \beta } =
\frac{i}{4}\left[ {\gamma ^\alpha ,\gamma ^\beta } \right]$, $\gamma
^\mu $ matrices satisfy $\left\{ {\gamma ^\mu ,\gamma ^\nu }
\right\} = 2g^{\mu \nu }I$. To obtain the solution of the Dirac equation, we should
choose $\gamma ^\mu $ matrices firstly. In this paper, we set
\begin{equation}
\gamma ^{\mu}=\left( -if^{-\frac{1}{2}}\sigma ^{2},f^{\frac{1}{2}}\sigma ^{1},\frac{1}{r}\sigma ^{3}\right) ,
\end{equation}%
in which $\sigma ^\mu $ are the Pauli sigma matrix
\begin{equation}
\sigma ^{1}=\left(
\begin{array}{cc}
0 & 1 \\
1 & 0
\end{array}%
\right) ,~\sigma ^{2}=\left(
\begin{array}{cc}
0 & -i \\
i & 0%
\end{array}%
\right) ,~\sigma ^{3}=\left(
\begin{array}{cc}
1 & 0 \\
0 & -1%
\end{array}
\right).
\end{equation}%
For a fermion with spin $1/2$,  the wave function have spin up state and spin
down state. In this paper, we only investigate the
spin up case for the case of spin down  is similar.
We use the ansatz for the two-component spinor $\psi $ as
\begin{equation}~~~~~~~~~~~~~~~
 \psi  = \left(
{{\begin{array}{*{20}c}
 {A\left( {t,r,\phi } \right)} \hfill \\
 {B\left( {t,r,\phi } \right)} \hfill \\
\end{array} }} \right)\exp \left( {\frac{i}{\hbar }I \left(
{t,r,\phi } \right)} \right).~~~~~~~~~~~~\label{wave0}
\end{equation}
Inserting Eq.(\ref{wave0})  into   Eq.(\ref{dirac}),
we have the following two simplified equations%
\begin{equation}
A\left( \mu +\frac{1}{r}\partial _{\phi }I \right) +B%
\left[ \sqrt{f}\partial _{r}I-\left( \frac{1}{\sqrt{f%
}}\partial _{t}I-\frac{1}{\sqrt{f}}e A_t \right) \right] =0,
\end{equation}%
\begin{equation}
A\left[ \sqrt{f}\partial _{r}I+\left( \frac{1}{\sqrt{%
f}}\partial _{t}I -\frac{1}{\sqrt{f}}e A_t \right) \right] +B\left( \mu -\frac{1}{r}\partial _{\phi
}I\right) =0.
\end{equation}%
These two equations have
a non-trivial solution for $A$ and $B$ if and only if the determinant of
coefficient matrix vanishes, which implies
\begin{equation}
\frac{1}{r^{2}}\left( \partial _{\phi }I
\right) ^{2}-\mu ^{2}+\left( \sqrt{f}\partial _{r}I\right) ^{2}-\left( \frac{1}{\sqrt{f}}\partial _{t}I -\frac{1}{\sqrt{f}}e A_t \right) ^{2}=0.  \label{wave}
\end{equation}
There are two Killing vectors
in the charged BTZ spacetime, so we can make the separation of variables for
$I\left( t,r,\phi \right) $ as
\begin{equation} \label{aI}
I=-\omega t+L\phi +I\left( r\right) +K,
\end{equation}%
where $\omega $ and $L$ are fermion's energy and angular momentum
respectively, and $K$ is a complex constant. Putting Eq.(\ref{aI}) into Eq.(\ref{wave}), we obtain
\begin{equation}
\partial _{r}I\left( r\right) =\pm \frac{1}{f}\sqrt{\left( \omega +eA_t \right) ^{2}+f\left( \mu ^{2}-\frac{L^{2}}{r^{2}}%
\right) }.
\end{equation}%
We are interested in the radial momentum of the particle $p^r\equiv g^{rr}p_r=g^{rr}\partial _{r}I\left( r\right)$. In addition,  we want to investigate the thermodynamics, so we will focus on the near horizon region. In this case, we get
\begin{equation}\label{emr}
\omega= |p^r_+|-eA_t(r_+),
\end{equation}
which is obviously the same as that of the scalar particles \cite{Gwak:2015fsa}. It should be stressed that a positive sign should be endowed in front of the $|p^r_+|$ term.  This choice is to assure that the signs in front of $\omega$ and $p^r_+$ are the same and positive in the positive flow of time.


\section{Thermodynamics and weak  cosmic  censorship  conjecture in the normal phase space }
\label{3}

The electrostatic potential difference between the black hole horizon and the infinity is
\begin{equation}
\Phi=-2 Q \log \left(\frac{r_+}{l}\right),
\end{equation}
in which $r_+$ is the event horizon of the black hole, which is determined by $f(r_+)=0$.
Based on the definition of  surface gravity, the Hawking temperature  can be expressed  as
\begin{equation}
T=\frac{r_+}{2 \pi  l^2}-\frac{2 Q^2}{ \pi  r_+}.\label{eq13}
 \end{equation}
For the three dimensional BTZ black hole, the black hole entropy can be expressed as
\begin{equation}
S=\frac{1}{2}\pi r_+.  \label{eq15}
 \end{equation}
In addition, with Eq.(\ref{frQ}),  and Eq.(\ref{mass}), the mass of the BTZ black hole can be expressed as
\begin{equation}
M=\frac{-8 l^2 Q^2 \log\left(\frac{r_+}{l}\right)+r_+^2}{8 l^2}.
 \end{equation}

As a
 charged fermion is absorbed by the black hole, the variation of the internal energy and charge  of the black hole satisfy
\be
\omega=dM, e=dQ,
\ee
in which the energy conservation and charge conservation  have been imposed. In this case,  Eq.(\ref{emr}) can be rewritten as
\be \label{emr1}
dM=\Phi dQ+p^r_+.
\ee
The absorbed fermions will change the configurations of the black holes. And there is a shift for the horizon of the black hole, labeled as $dr_{+}$. In the new
horizon, there is also a relation, $f(r_++dr_+)=0$. In other words,  the change of the horizon    should satisfy
\bea \label{df}
df_+&=&f(r_++dr_+)-f(r_+) \nonumber \\
&=&\frac{\partial f_+}{\partial M}dM+\frac{\partial f_+}{\partial Q}dQ+\frac{\partial f_+}{\partial r_+}dr_+=0.~~~~
\ena
Note that here, $q$ and $m$ in Eq.(\ref{frQ}) have been substituted by $Q$ and $M$ in Eq.(\ref{charge}) and Eq.(\ref{mass}).
Inserting Eq.(\ref{emr1}) into Eq.(\ref{df}), we can delete $dM$. Interestingly $dQ$ is eliminated meanwhile. Solving this equation,  we can get  $dr_+$  directly, which is
 \be\label{dr}
dr_+=-\frac{4 l^2 p^r_+  r_+}{4 l^2 Q^2-r_+^2}.
\ee
Based on Eq.(\ref{dr}),  we can get   the variation of entropy by making use of Eq.(\ref{eq15}), that is
\begin{equation}\label{ds}
dS=-\frac{\pi}{2}\frac{4 l^2p^r_+ r_+}{4 l^2 Q^2-r_+^2}.
\end{equation}
With Eqs.(\ref{eq13}) and (\ref{ds}), we find there is a relation
\be
T dS=p^r_+.
\ee
In this case, the internal energy in Eq.(\ref{emr1}) can be rewritten as
\begin{equation} \label{eq:dm1}
dM=TdS+\Phi dQ,
\end{equation}
which is  the first law of   black hole thermodynamics. That is to say, as a fermion drops into the black hole, the first law
is valid in the normal phase space.

Next we concentrate on studying the second law of  thermodynamics, which states that the entropy of the black holes never decrease in the clockwise direction. As a fermion is absorbed by the black hole, the entropy of the black hole increases according to the second law of the thermodynamics. We will employ Eq.(\ref{ds}) to check whether this is true.

For the extremal  black holes, the temperature vanishes at the horizon for the inner horizons and outer horizons are coincident.
  With Eq.(\ref{eq13}), we can get the mass of the extremal black hole and substitute it  into Eq.(\ref{ds}), we find
\be\label{dse}
dS_{extreme}=\infty.~~~~~~
\ee
The divergence of $dS$ implies the second law for the extremal black hole is meaningless  since the thermodynamic system is  a zero temperature system.

For the non-extremal  black holes, their temperature are larger then zero, which implies
\be\label{eq:variables00013}
r_+^2>4 l^2 Q^2,
\ee
where we have used Eq.(\ref{eq13}). In this case, $dS$ in  Eq.(\ref{ds}) is positive.  The second law of thermodynamics is valid therefore.

In the normal phase space, we also can check the validity of  the weak
cosmic censorship conjecture, which states that the singularity of a spacetime can not be observed for an  observer located at future null infinity. In other words, singularities need to be hidden  by the event horizon for a black hole. So an event horizon should exist to assure the validity of the  weak cosmic censorship conjecture. As a fermion is absorbed by a black hole, we intend to check whether there is an event horizon. That is, whether the equation  $f(r)=0$ has solutions.

 For the BTZ black holes, there is a minimum value  for $f(r)$ with the radial coordinate $r_{m}$. When $f(r_m)>0$, there is not a horizon while when $f(r_m)\leq0$, there are horizons always.   At $r_{m}$,
 the following relations \cite{Zeng:2019aao,He:2019fti,Zeng:2019hux}
\bea \label{condition}
&f(r)|_{r=r_{m}}\equiv f_{m}=\epsilon\leq 0, \nonumber\\
&\partial_{r}f(r)|_{r=r_{m}}\equiv f'_{m}=0,\label{condition}
\ena
should satisfy. For the extremal black holes, $\epsilon=0$,  $r_+$ and and $r_m$ are coincident. For the near extreme black holes, $\epsilon$ is a small quantity, $r_{m}$ is distributed between the inner horizon and outer horizon.  As a fermion drops into the black hole, the mass and charge of the black hole  change into $M+dM, Q+dQ$ respectively. Correspondingly, the locations of
the minimum value and event horizon change into $ r_{m}+dr_{m}$, $r_{+}+dr_{+}$. There is also a shift for $f(r)$, which can be written as
\bea \label{dfm}
df_{m}=f(r_{m}+dr_{m})-f_{m}=\left(\frac{\partial f_{m}}{\partial M}dM+\frac{\partial f_{m}}{\partial Q}dQ\right),
\ena
where we have used $f'_{m}=0$ in Eq.(\ref{condition}). We first discuss the extremal black holes, for which the horizons are located at $r_m$. In this case, Eq.(\ref{emr1}) can be used. Inserting Eq.(\ref{emr1}) into Eq.(\ref{dfm}), we find $dQ$ is deleted meanwhile. In this case, Eq.(\ref{dfm}) can be simplified lastly as
\be
 df_{m}=-8p^r_+. \label{eqc6}
\ee
This result shows that $f(r_{m}+dr_{m})$  is smaller than $f(r_{m})$  as a charged  fermion is absorbed by the black hole.

For the near-extremal black holes, Eq.(\ref{emr1})  is not valid at $r_m$ for it holds only at the horizon. With the condition $r_{+}=r_m+\delta$, we can expand Eq.(\ref{emr1}) at
$r_m$, which leads to
\be
dM=p^r_+ -2  {dQ} Q\log\left(\frac{r}{R}\right)-\frac{2  Q{dQ} }{r}\delta+O(\delta )^2. \label{eqc66666}
\ee
Substituting Eq.(\ref{eqc66666}) into Eq.(\ref{df}), we get lastly
\be
df_{m}=-8p^r_+  +\frac{32  {dQ}}{l}\delta+O(\delta)^2. \label{eqc666}
\ee
 Because $\delta$ is a small quantity while $l$ is a large quantity relatively, the last two terms can be neglected approximately.  In this case, Eq.(\ref{eqc666}) takes the same form as Eq.(\ref{eqc6}), indicating  that the weak
cosmic censorship conjecture is also valid for the near-extremal black holes.

It should be stressed that the  second term in Eq.(\ref{eqc666}) is small for we compare it with $8p^r_+$. In fact, the higher order corrections are important for us to discuss the weak
cosmic censorship conjecture, however in our method, we find they can be neglected after calculation strictly for the dominant term is too large.

\section{Thermodynamics and weak  cosmic  censorship  conjecture in the extended  phase space}

\label{4}

To make the charged BTZ black holes satisfy the Reverse Isoperimetric Inequality, a new thermodynamic parameter $R$ was introduced in the first law \cite{Frassino:2015oca}, that is
\be
dM=TdS+VdP+\Phi dQ+KdR,\label{9}
\ee
where
\bea
M&=&\frac{r_+^2-8 l^2 Q^2 \log \left(\frac{r_+}{R}\right)}{8 l^2},\label{10}
\\
P&=&-\frac{\Lambda}{8 \pi}=\frac{1}{8 \pi l^2},
\\
V&=&\left(\frac{\partial M}{\partial P}\right)_{S,Q,R}=\pi r_+^2,\label{11}
\\
\Phi&=&\left(\frac{\partial M}{\partial Q}\right)_{S,Q,R}=-2 Q \log \left(\frac{r}{R}\right),\label{12}
\\
K&=&\left(\frac{\partial M}{\partial R}\right)_{S,Q,P}=Q^2/R,\label{133}
\ena
in which $R$ is the renormalization length scale, and $K$, which  is the conjugate  of $R$,   is   the renormalized energy. Note that  here the value of $K$ is different from  that in \cite{Frassino:2015oca}. The reason stems from   the definition of the electric charge $Q$. In fact, with the Gauss law, we have noted that the  charge parameter $q$ is not the electric charge of the black hole, which has been shown in  Eq.(\ref{charge}).

From   Eq.(\ref{11}), we know that in this framework, the volume recovers to the standard definition of the thermodynamic volume \cite{Frassino:2015oca}, which is more reasonable. We are going to explore whether the first law in Eq.(\ref{9}) can be obtained by considering a charged fermion absorbtion.

In the extended, the pressure $P$  and the renormalization length scale $R$ are also state parameters of the thermodynamic system, as a fermion is absorbed by the black hole,   the pressure   and the renormalization length scale  will also change besides the mass, charge and entropy. In this thermodynamic system, the mass $M$ is not the internal energy but the  enthalpy, which relates to the internal energy   as \cite{Frassino:2015oca}
\be
M=U+PV+KR. \label{eq119}
\ee
As a charged fermion drops into the black hole, the energy and charge are supposed to be conserved. Namely the energy and charge of the fermion equal to the varied energy and charge of the black hole, which implies
\be
\omega=dU=d(M-PV-KR),\quad e=dQ,
\ee
 The energy in Eq.(\ref{emr}) changes correspondingly into
\be \label{emr2}
dU=\Phi dQ+p^r_+.
\ee
Considering the backreaction, the absorbed fermions will change  the location of the event horizon of the black hole. However, the horizon is determined by the equation $f(r)=0$ always as stressed in section 3. In the extended phase space, for the AdS radius $l$ and renormalization length $R$  are variables, the shift of function  $f(r)$  can be expressed as
\be \label{eq:function05}
df_+=\frac{\partial f_ +}{\partial M}dM+\frac{\partial f_ +}{\partial Q}dQ+\frac{\partial f_ +}{\partial l}dl+\frac{\partial f_ +}{\partial r_ {+}}dr_++\frac{\partial f_+}{\partial R}dR=0.
\ee
In addition, with Eq.(\ref{eq119}), Eq.(\ref{emr2}) can be expressed as
\be \label{eq:dispersion03}
dM-d(PV+KR)=\Phi dQ+p^r_+.
\ee
From Eq.(\ref{eq:function05}), we can obtain $dl$. Substituting  $dl$ into Eq.(\ref{eq:dispersion03}), we can delete it directly. Interestingly, $dQ$, $dR$, and $dM$ are also eliminated at the same time. In this case, there is only a  relation between $p^r_+$  and  $dr_{+}$,  which is
 \be \label{eq:variables02}
dr_ {+}=-\frac{r_+ (p^r_+ +{dK} R)}{Q^2}.
\ee
Based on Eq.(\ref{eq:variables02}), the variations of entropy and volume of the black hole can be expressed as
\be\label{eq:variables001}
dS=-\frac{\pi  r_+ (p^r_+ +{dK} R)}{2 Q^2},
\ee
\be\label{eq:variables002}
dV=-\frac{2 \pi  r_+^2 (p^r_++{dK} R)}{Q^2}.
\ee
With Eq.(\ref{eq:variables001}) and  Eq.(\ref{eq:variables002}), we find
\be \label{pr}
T dS-PdV-R dK=p^r_+.
\ee
The internal energy in Eq.(\ref{emr2}) thus would change into
\be  \label{du}
dU=\Phi dQ + T dS-PdV-R dK.
\ee
 Moreover, from Eq.(\ref{eq119}), we can get
\be  \label{dm1}
dM=d U+P dV+VdP+KdR+RdK.
\ee
Substituting Eq.(\ref{dm1}) into Eq.(\ref{du}), we find
\be
dM=TdS+\Phi dQ+VdP+KdR,
\ee
which is consistent with that in Eq.(\ref{9}). That is, as a charged fermion is absorbed by the black hole, the first law of thermodynamics holds in the extended phase space.

With Eq.(\ref{eq:variables001}), we also can check  the second law of thermodynamics in the extended phase space.
It should be stressed that there is a term $dK$ in  Eq.(\ref{eq:variables001}), which is the variation of the
renormalized energy.  According to Eq.(\ref{133}), $dK$  is the function of  $dR, dQ$. However, the existence of  $dQ, dR$ would affect the definition of $\Phi$ and  $K$ respectively and further violate the first law of thermodynamics.
The satisfaction of the first law of thermodynamics is a necessary condition to discuss the second law of thermodynamics under a particle absorption. So, $dK$ can not be expressed as a linear relation  of $dR, dQ$ though we do not know the  mechanism  for we know little about the
renormalized energy in the extended phase space. In this paper, we will treat the variation of the renormalized energy as an independent  quantity and do not care about its form.

From Eq.(\ref{eq:variables001}), we know  the variation of the entropy depends on the variation of the renormalized energy. For the case, $dK>-p^r_+/R$,  $dS$ is negative and for the case $dK<-p^r_+/R$, $dS$ is positive. In other words, the second law is violated for the case
$dK>-p^r_+/R$, and valid for the case  $dK<-p^r_+/R$. Moreover, for $dK=-p^r_+/R$, $dS=0$, indicating that the horizons of the black holes will not change  as a charged fermions is absorbed.

We also can discuss the   weak
cosmic censorship conjecture in the extended phase space with the condition in Eq.(\ref{condition}).
Because of  the backreaction, the mass $M$, charge $Q$,  renormalization length $R$, and AdS radius $l$ of the black hole will change into $(M+dM, Q+dQ, R+dR, l+dl)$ as a charged fermion
drops into the black hole. Correspondingly, the locations of
the minimum value, event horizon, AdS radius, and renormalization length will change into $ r_{m}+dr_{m}$, $ r_{+}+dr_{+}$, $ l+dl$, $ R+dR$.
In this case, the shift of $f(r)$ can be written as
\be  \label{eeqc1}
df(r_{m})=\left(\frac{\partial f_{m}}{\partial M}dM+\frac{\partial f_{m}}{\partial Q}dQ+\frac{\partial f_{m}}{\partial l}dl+\frac{\partial f_{m}}{\partial R}dR\right),
\ee
where we have used $f'_{m}=0$ in Eq.(\ref{condition}).
Next, we focus on  finding the last result of Eq.(\ref{eeqc1}). For the extremal black holes, the horizons are located at $r_m$.
The energy relation in Eq.(\ref{eq:dispersion03}) is valid. Substituting Eq.(\ref{eq:dispersion03}) into  Eq.(\ref{eeqc1}), we find
\be
df(r_{m})
=-8 p^r_+- 8 dK R-\frac{2  r_{m}{dr_{m}}}{l^2}, \label{eqc1110}
\ee
Substituting Eq.(\ref{pr}) into  Eq.(\ref{eqc1110}), we find
\be \label{dfe}
df(r_{m})=0.
\ee
 That is, as fermions drop into the extremal BTZ  black holes, the black holes  stay at their initial states so that their configurations will not be changed. This result is quite
different from that in the normal phase space  where the extremal black holes will evolve into the non-extremal black holes by the absorption.

For the near-extremal black hole,  Eq.(\ref{eq:dispersion03}) is not valid. But we can expand it near the lowest point with $r_+=r_m+\delta$. It should be stressed that $p^r_+$ should also be expanded for it is also a function of the horizon $r_+$. To the first order, we get
\bea
dM&=&-\frac{ r_m^2}{4 l^3}{dl}-2 Q \log \left(\frac{r_m}{R}\right){dQ}+\frac{ r_m {dr}}{4 l^2}-\frac{ Q^2{dr}}{r_m}+\frac{ Q^2{dR}}{R}\nonumber\\
&+&  \left(-\frac{ r_m{dl}}{2 l^3}-\frac{2 {dQ}Q}{r_m}+\frac{{dr}}{4 l^2}+\frac{ Q^2{dr} }{r_m^2}\right)\delta+O(\delta)^2, \label{eqc111}
\ena
Substituting Eq.(\ref{eqc111}) into Eq.(\ref{eeeqc2}), we can get lastly
\bea\label{dfrm}
df(r_{m})
&=& \left(\frac{8 Q^2}{r_m}-\frac{2 r_m}{l^2}\right){dr_m} \nonumber \\
&+& \left(\frac{4  r_m{dl}}{l^3}+\frac{16  Q{dQ}}{r_m}-\frac{2 {dr}}{l^2}-\frac{8 Q^2{dr} }{r_m^2}\right)\delta \nonumber \\
&+&O(\delta)^2,
\ena
In addition,  at the $r_m+dr_m$, there is also a relation
\be
\partial_{r} f(r)|_{r=r_m+dr_m}
=f'_{m}+df'_{m}=0,
\ee
which implies
\be \label{eeeqc2}
df'_{m}=\frac{\partial f'_{m}}{\partial Q}dQ+\frac{\partial f'_{m}}{\partial l}dl+\frac{\partial f'_{m}}{\partial r_{m}}dr_{m}=0.
\ee
Solving this equation, we obtain
\be \label{dl}
dl=\frac{l \left(-8 l^2 Q r_{m} {dQ}+4 l^2 Q^2{dr_{m}}+ r^2{dr_{m}}\right)}{2 r_{m}^3}.
\ee
Based on the condition
$f'_{m}=0$ in Eq.(\ref{condition}), we can get
\be \label{eqc3}
l=\frac{r_{m}}{2 Q}.
\ee
Substituting Eq.(\ref{eqc3}) and Eq.(\ref{dl}) into Eq.(\ref{dfrm}), we  find
\bea\label{dfrm1}
df(r_{m})=
O(\delta)^2,
\ena
which  shows that  the  near-extremal black holes are also stable. This result is consistent with the extremal black holes in Eq.(\ref{dfe}). So we can conclude that the  weak
cosmic censorship conjecture holds for both the extremal and near-extremal black holes in the extended phase space for the configurations of the black holes are not changed as fermions are adsorbed.

\section{Conclusions}
\label{5}

In the normal phase space and extended phase space, the laws of thermodynamics and  weak cosmic censorship conjecture  in the  BTZ black holes were checked under a charged fermion absorbtion. We investigated firstly the motion of a fermion via the Dirac equation and obtained  a relation between the energy and momentum near the horizon. With this relation, the first law was reproduced in the normal phase space. By studying the variation of the entropy, we also checked the second law of thermodynamics and found that for both the extremal black holes and near-extremal black holes, the second law was valid in the normal phase space for the variation of the entropy is positive. The  weak cosmic censorship censorship for  the extremal black holes and near-extremal black holes were checked too. We found  that  the metric function  which determine the locations of the horizons    moved  with the same scale, $-8p^r_+$, implying that there are always horizon to hidden the singularity  so that the weak cosmic censorship  are  valid for both cases.

With similar strategy, the thermodynamic laws and  weak cosmic censorship conjecture were investigated in the extended phase space further.
We found that the first law of thermodynamics  was still  valid,  but the validity of the second law    depended on the variation of the renormalization energy
$dK$.  For the case $dK>-p^r_+/R$, the second law is violated and  for the case $dK\leq-p^r_+/R$, the second law is valid.  Though the  weak cosmic censorship conjecture are valid in both the normal and extended phase space for the extremal and near-extremal  black holes, their final states are different  after absorbtion. The extremal and near-extremal black holes will evolve into non-extremal black holes in the normal phase space, while they are stable in the extended phase space.

In \cite{Gwak:2015fsa},   laws of thermodynamics and the weak
cosmic censorship conjecture of the BTZ black holes have been investigated. Different from \cite{Gwak:2015fsa}, in this paper, the absorbed particles are fermions. In addition, laws of  thermodynamics and the weak
cosmic censorship conjecture were discussed not only in the normal phase space  but also in the extended phase space in this paper while  \cite{Gwak:2015fsa} only investigated the case of normal phase space.

\end{multicols}

\vspace{10mm}

\vspace{-1mm}
\centerline{\rule{80mm}{0.1pt}}
\vspace{2mm}

\begin{multicols}{2}

\end{multicols}

\clearpage
\end{CJK*}
\end{document}